\documentclass[journal,onecolumn]{IEEEtran}
\usepackage{amsmath, epsfig, cite}
\usepackage{amssymb}
\usepackage{amsfonts}
\usepackage{latexsym}
\usepackage{longtable}
\usepackage{tabularx}
\usepackage{color}

\newtheorem{theorem}{Theorem}

\title{A general private information retrieval scheme for MDS coded databases with colluding servers}

\author{Yiwei Zhang and Gennian Ge
\thanks{The research of G. Ge was supported by the National Natural Science Foundation of China under Grant Nos. 11431003 and 61571310, Beijing Hundreds of Leading Talents Training Project of Science and Technology, and Beijing Municipal Natural Science Foundation.}
\thanks{Y. Zhang is with the School of Mathematical Sciences, Capital Normal University, Beijing 100048, China (email: rexzyw@163.com).}
\thanks{G. Ge is with the School of Mathematical Sciences, Capital Normal University, Beijing 100048, China. He is also with Beijing Center for Mathematics and Information Interdisciplinary Sciences, Beijing 100048, China (e-mail: gnge@zju.edu.cn).}
}

\begin{document}

\date{}\maketitle

\begin{abstract}

The problem of private information retrieval gets renewed attentions in recent years due to its information-theoretic reformulation and applications in distributed storage systems. PIR capacity is the maximal number of bits privately retrieved per one bit of downloaded bit. The capacity has been fully solved for some degenerating cases. For a general case where the database is both coded and colluded, the exact capacity remains unknown. We build a general private information retrieval scheme for MDS coded databases with colluding servers. Our scheme achieves the rate $(1+R+R^2+\cdots+R^{M-1})$, where $R=1-\frac{{{N-T}\choose K}}{{N\choose K}}$. Compared to existing PIR schemes, our scheme performs better for a certain range of parameters and is suitable for any underlying MDS code used in the distributed storage system.
\end{abstract}


\section{Introduction} \label{SecIntro}

The classic model of a private information retrieval (PIR) scheme, first introduced by Chor et al. \cite{Chor1,Chor2}, allows a user to make queries to several servers storing the same database and retrieve a certain bit of the database via the feedbacks, without revealing the identity of the specific bit to any single server. Ever since its introduction, PIR has received a lot of attention from the computer science community and cryptography community. The main research objective is to minimize the total communication cost for retrieving a single bit from an $n$-bit database, where the communication cost includes both the upload cost (the queries sent to the servers) and the download cost (the feedbacks from the servers). A series of works have managed to reduce the communication cost, see for example, \cite{Beimel,Efremenko,Dvir,Yekhanin,YekhaninSurvey} and the references therein.

While the classic model of PIR has been extensively studied, PIR gets renewed attentions in recent years, mainly due to the following two aspects. First, PIR could be reformulated from an information-theoretic point of view. Instead of retrieving a single bit from an $n$-bit database, the information-theoretic PIR problem considers retrieving a single file from an $n$-file database, where the length of each file could be arbitrarily large. The research objective is then minimizing the total communication cost per unit of retrieved data. It is further noticed in \cite{Chan} that in this information-theoretic model the upload cost could be neglected with respect to the download cost. Second, original PIR schemes are restricted to replicated-based databases, i.e., every server stores the whole database. With the development of distributed storage system (DSS), it is natural to consider designing PIR schemes for coded databases \cite{Augot,Shah,Chan,Blackburn}.

Currently there are mainly two lines of research regarding the information-theoretic and DSS-oriented PIR problems. One originates from \cite{FazeliConfversion,FazeliFullversion}, where PIR codes and subsequently PIR array codes are introduced. These codes characterize how to design distributed storage systems that may emulate any classic $k$-server PIR scheme, with relatively a smaller storage overhead. Researches along this line include, for example, \cite{Asi,BlackburnPIRcodes,Rao,Vajha,Zhang}.

The other line, which is the focus of this paper, is the fundamental limits on PIR capacity for a general PIR scheme defined as follows. Assume we have a coded database containing $N$ servers storing $M$ files. Each file is stored independently via the same arbitrary $(N,K)$-MDS code. Any $T$ out of the $N$ servers may collude in attacking on the privacy of the user. We call the PIR scheme for such a system as an $(N,K,T;M)$-scheme. PIR rate is defined as the number of bits of messages that could be privately retrieved per one bit of downloaded message. The supremum of all achievable rates is called the PIR capacity, denoted by $C=C(N,K,T;M)$. In the pioneering work \cite{Sun} of Sun and Jafar, the exact PIR capacity for the case $K=T=1$ is completely solved. Later \cite{SunColluded} deals with the case when $K=1$ and $T$ is arbitrary, i.e., colluded databases. Then \cite{BanawanCoded} deals with the case when $T=1$ and $K$ is arbitrary, i.e., coded databases. The PIR capacity for all the three models above are of the same form:
\begin{equation}
C=(1+R+R^2+\cdots+R^{M-1})^{-1},   \label{CapacityGeneralForm}
\end{equation}
where $C$ equals $1/N$, $T/N$ or $K/N$ accordingly.

For the databases both coded and colluded, i.e., $K\ge2$ and $T\ge2$, determining the exact value of PIR capacity is difficult. In \cite{TajeddineConfversion,TajeddineFullversion} a PIR scheme for the case $T+K\le N$ is posed and it has PIR rate $\frac{1}{K+T}$. In \cite{Freij-Hollanti} Freij-Hollanti et al. introduce a PIR scheme with rate $\frac{N-K-T+1}{N}$ for any $T+K\le N$. So far this result is the best among known PIR schemes for a large range of parameters. However, one limitation of this result is that it depends on some special properties of the underlying MDS code used in the coded database. For example, the scheme in \cite{Freij-Hollanti} holds when the storage code is a generalized Reed-Solomon code. But for any other arbitrary storage code, it is uncertain whether similar results hold. Based on the intuition from existing explicit expressions shown in (\ref{CapacityGeneralForm}), in \cite{Freij-Hollanti} Freij-Hollanti et al. further conjecture that if $T+K\le N$, then the optimal PIR capacity is given by
\begin{equation}
C=(1+R+R^2+\cdots+R^{M-1})^{-1},\text{~~~~}R=\frac{T+K-1}{N}.  \label{FGHKConjecture}
\end{equation}

In a very recent work \cite{SunNew}, the conjecture above is disproved by analyzing the case with parameters $N=4$, $T=2$, $K=2$ and $M=2$. The conjectured capacity is $4/7$ while \cite{SunNew} poses a scheme with rate $3/5>4/7$. A further counterexample is a complete analysis of the case with parameters $M=2$ and $K=N-1$ \cite{SunNew}. For the other parameters, whether the conjecture (\ref{FGHKConjecture}) holds or not is an open problem on both sides, i.e., we neither have an explicit PIR scheme achieving this capacity nor have a good upper bound of the optimal capacity (which may require a subtle analysis using information inequalities).

In this paper, our main contribution is a general PIR scheme for the case $T+K\le N$ with rate given by
\begin{equation}
C=(1+R+R^2+\cdots+R^{M-1})^{-1},\text{~~~~}R=1-\frac{{{N-T}\choose K}}{{N\choose K}}.  \label{OurRate}
\end{equation}

Several remarks on this PIR scheme are as follows.

$\bullet$ For a certain range of parameters, our scheme has better rate than any other existing scheme.

$\bullet$ For the degenerating cases, i.e., either $K=1$ or $T=1$, our rate agrees with previous results. For the non-degenerating cases, note that our rate in (\ref{OurRate}) is strictly smaller than the conjectured value (\ref{FGHKConjecture}). So our scheme offers a lower bound of the optimal capacity and does not provide more counterexamples for the conjecture by Freij-Hollanti et al.

$\bullet$ When the number of files $M$ is large, the scheme by Freij-Hollanti et al. in \cite{Freij-Hollanti} is better than ours. However, as already mentioned, their results rely on certain properties of the underlying storage code while ours is a general scheme suitable for any $(N,K)$-MDS code.

The rest of the paper is organized as follows. In Section \ref{SecModel}, we introduce the general model of the PIR problem. In Section \ref{SecScheme} we present our main contribution and analyze its rate. In Section \ref{SecComparison} we make a comparison of some known PIR schemes. Section \ref{SecConclusion} concludes the paper.

\section{Problem Statement}\label{SecModel}

In this section we introduce the general model of the PIR problem. Basically the statement follows the same way as shown in \cite{BanawanCoded,Sun,SunColluded,SunNew}.

The general model considers a distributed storage system consisting of $N$ servers. The system stores $M$ files, denoted as $W^{[1]},W^{[2]},\dots,W^{[M]}\in\mathbb{F}_q^{L\times K}$, i.e., each file is of length $LK$ and represented in a matrix of size $L\times K$. The files are independent and identically distributed with
\begin{align}
H(W^{[i]})&=LK,~~~i\in\{1,2,\dots,M\},\\
H(W^{[1]},W^{[2]},\dots,W^{[M]})&=MLK.
\end{align}

Denote the $j$th row of the file $W^{[i]}$ as $\mathbf{w}_j^{[i]}\in\mathbb{F}_q^{K}$, $1\le j \le L$. Each file is stored in the system via the same given $(N,K)$-MDS code. The generator matrix $\mathbf{G}\in \mathbb{F}_q^{K\times N}$ of the MDS code is denoted as
\begin{equation}
\mathbf{G}=\Big[ \mathbf{g_1}~~\mathbf{g_2}~~\cdots~~\mathbf{g_N} \Big]_{K\times N},
\end{equation}
and the MDS property means that any $K$ columns of $\mathbf{G}$ are linearly independent. For each $\mathbf{w}_j^{[i]}$, the $n$th server stores the coded bit $\mathbf{w}_j^{[i]}\mathbf{g_n}$. Thus, the whole contents $\mathbf{y}_n\in\mathbb{F}_q^{ML}$ stored on the $n$th server are the concatenated projections of all the files $\{W^{[1]},W^{[2]},\dots,W^{[M]}\}$ on the encoding vector $\mathbf{g_n}$, i.e.,
\begin{align}
  \mathbf{y}_n&=\left(
                  \begin{array}{c}
                    W^{[1]} \\
                    \vdots \\
                    W^{[M]} \\
                  \end{array}
                \right)\mathbf{g_n}\\
  &=\Big[ \mathbf{w}_1^{[1]}\mathbf{g_n}~~\cdots~~\mathbf{w}_L^{[1]}\mathbf{g_n}~~\mathbf{w}_1^{[2]}\mathbf{g_n}~~\cdots~~\mathbf{w}_L^{[2]}\mathbf{g_n}~~\cdots~~\mathbf{w}_1^{[M]}\mathbf{g_n}~~\cdots~~\mathbf{w}_L^{[M]}\mathbf{g_n} \Big]^T.
\end{align}

Now assume that a user wants to retrieve an arbitrary file $W^{[i]}$. This is done by sending some queries to the servers and then getting the feedbacks. Let $\mathcal{F}$ denote a random variable generated by the user and unknown to any server. $\mathcal{F}$ represents the randomness of the user's strategy to generate the queries. Let $\mathcal{G}$ denote a random variable generated by the servers and also known to the user. $\mathcal{G}$ represents the randomness of the server's strategy to produce the feedbacks\footnote{The role of $\mathcal{G}$ deserves a special remark. This is first proposed by Sun and Jafar in \cite{SunNew}. In almost all previous PIR schemes we follow a question-and-answer format, i.e., for any query vector the server responds the corresponding projection of his contents onto the query. By bringing a strategy $\mathcal{G}$ into consideration, the server can perform some coding procedures before responding and thus could reduce the download cost. For more details please refer to \cite{SunNew} or the discussions in Section \ref{SecComparison}.}. Both $\mathcal{F}$ and $\mathcal{G}$ are generated independently of the files and the identity of the desired file, i.e.,
\begin{equation}
  H(\mathcal{F},\mathcal{G},i,W^{[1]},W^{[2]},\dots,W^{[M]})=H(\mathcal{F})+H(\mathcal{G})+H(i)+H(W^{[1]})+\cdots+H(W^{[M]}).
\end{equation}

Using his strategy $\mathcal{F}$, the user generates a set of queries $Q^{[i]}_n$ to the $n$th server, $1\le n \le N$. The queries are independent of the files, i.e.,
\begin{equation}
  I(Q^{[i]}_1,Q^{[i]}_2,\dots,Q^{[i]}_N;W^{[1]},W^{[2]},\dots,W^{[M]})=0.
\end{equation}

Upon receiving the query, the $n$-th server responds a feedback $A^{[i]}_n$, which is a deterministic function of the query $Q^{[i]}_n$, the strategy $\mathcal{G}$ and the data $\mathbf{y}_n$ (and therefore a deterministic function of the query $Q^{[i]}_n$, the strategy $\mathcal{G}$ and the files $\{W^{[1]},W^{[2]},\dots,W^{[M]}\}$), i.e.,
\begin{equation}
  H(A^{[i]}_n|Q^{[i]}_n,\mathcal{G},\mathbf{y}_n)=H(A^{[i]}_n|Q^{[i]}_n,\mathcal{G},W^{[1]},W^{[2]},\dots,W^{[M]})=0.
\end{equation}

The user retrieves his desired file based on all the queries and feedbacks, plus the knowledge of the strategies $\mathcal{F}$ and $\mathcal{G}$, i.e.,
\begin{equation}
  H(W^{[i]}|Q^{[i]}_1,Q^{[i]}_2,\dots,Q^{[i]}_N,A^{[i]}_1,A^{[i]}_2,\dots,A^{[i]}_N,\mathcal{F},\mathcal{G})=0.
\end{equation}

For any subset $\mathcal{T}$ of the servers, $|\mathcal{T}|=T$, let $Q^{[i]}_\mathcal{T}$ represent $\{Q^{[i]}_n,n\in\mathcal{T}\}$. Similarly we have the notation $A^{[i]}_\mathcal{T}$. The PIR scheme should ensure that this set of servers learns nothing about the identity of the retrieved file, i.e.,
\begin{equation}
  I(i;Q^{[i]}_\mathcal{T})=I(i;A^{[i]}_\mathcal{T})=0, ~\forall\mathcal{T}\subseteq\{1,2,\dots,N\},~|\mathcal{T}|=T.
\end{equation}

The rate for the PIR scheme is defined as the ratio of the size of the retrieved file to the total download cost, i.e.,
\begin{equation}
  \frac{H(W^{[i]})}{\sum_{n=1}^N H(A^{[i]}_n)},
\end{equation}
and the PIR capacity $C$ is the supremum of all achievable rates.

\section{A general PIR scheme}\label{SecScheme}

In this section we introduce our main result. We first demonstrate the basic idea via two explicit examples before analyzing the general scheme.

\subsection{$N=4$, $K=2$, $T=2$ and $M=2$}

Denote the two files by $U$ and $V$ and assume that each file is of length $72$ and represented in a matrix of size $36\times2$, i.e.,
\begin{equation}
  U=\Bigg(
                  \begin{array}{c}
                    \mathbf{u}_1 \\
                    \vdots \\
                    \mathbf{u}_{36} \\
                  \end{array}
                \Bigg)
  \hspace{3em}
  V=\Bigg(
                  \begin{array}{c}
                    \mathbf{v}_1 \\
                    \vdots \\
                    \mathbf{v}_{36} \\
                  \end{array}
                \Bigg)
\end{equation}
where $\mathbf{u}_i\in\mathbb{F}_q^{2}$ and $\mathbf{v}_i\in\mathbb{F}_q^{2}$, $1\le i \le 36$. Here $\mathbb{F}_q$ is a sufficiently large finite field\footnote{The only constraint on size of the field is to allow the existence of an MDS code used in the construction later.}. Let $U$ be the desired file.

Choose a random matrix $S_1\in\mathbb{F}_q^{36\times36}$ uniformly from all the $36\times36$ full rank matrices over $\mathbb{F}_q$. Construct a list of {\it atoms}\footnote{Here we use the word ``atom" to represent a query only related to one single file. The query related to more than one file will then be consisted of several atoms. This is for reducing too frequent appearances of the word ``query", which is annoying and may lead to confusion.} $a_{[1:36]}=S_1U$. Note that this is only a formal expression and should be understood as follows. For example, if the first row of $S_1$ is the vector $(p_1,p_2,\dots,p_{36})$, then $a_1$ represents the linear combination $a_1=p_1\mathbf{u}_1+p_2\mathbf{u}_2+\cdots+p_{36}\mathbf{u}_{36}$.

Choose a random matrix $S_2\in\mathbb{F}_q^{36\times36}$ uniformly from all the $36\times36$ full rank matrices over $\mathbb{F}_q$. Suppose there exists a $(36,30)$-MDS code and the transpose of its generator matrix is denoted as $\text{MDS}_{36\times30}$. Select the first 30 rows of $S_2$, denoted as $S_2[(1:30),:]$. Construct a list of atoms $b_{[1:36]}=\text{MDS}_{36\times30}S_2[(1:30),:]V$. Note that this is only a formal expression and should be understood as follows. For example, if the first row of $\text{MDS}_{36\times30}S_2[(1:30),:]$ is the vector $(q_1,q_2,\dots,q_{36})$, then $b_1$ represents the linear combination $b_1=q_1\mathbf{v}_1+q_2\mathbf{v}_2+\cdots+q_{36}\mathbf{v}_{36}$.

The atoms $a_{[1:36]}$ and $b_{[1:36]}$ will form the queries for the servers. The queries for the servers are divided into eleven blocks. In ten blocks each query is only a single atom and in one block $\Lambda^{a+b}$ each query is a combination of two atoms. The query structure is as follows.

$$\Lambda^x_{\lambda}:\Bigg\{\begin{array}{cccc}
  \text{Server I} & \text{Server II} & \text{Server III} & \text{Server IV}\\\hline
  x_{6\lambda+1} & x_{6\lambda+1} & x_{6\lambda+2} & x_{6\lambda+2} \\
  x_{6\lambda+3} & x_{6\lambda+4} & x_{6\lambda+3} & x_{6\lambda+4} \\
  x_{6\lambda+5} & x_{6\lambda+6} & x_{6\lambda+6} & x_{6\lambda+5} \\
\end{array}\Bigg\}$$
for $\lambda\in\{1,2,3,4,5\}$, $x\in\{a,b\}$ and
$$\Lambda^{a+b}:\Bigg\{\begin{array}{cccc}
  \text{Server I} & \text{Server II} & \text{Server III} & \text{Server IV}\\\hline
  a_{1}+b_{1} & a_{1}+b_{1} & a_{2}+b_{2} & a_{2}+b_{2} \\
  a_{3}+b_{3} & a_{4}+b_{4} & a_{3}+b_{3} & a_{4}+b_{4} \\
  a_{5}+b_{5} & a_{6}+b_{6} & a_{6}+b_{6} & a_{5}+b_{5} \\
\end{array}\Bigg\}.$$

Recall that the $n$-th server, $1\le n \le 4$, has stored
$\mathbf{y}_n=\left(
                  \begin{array}{c}
                    U \\
                    V \\
                  \end{array}
                \right)\mathbf{g_n}$, where $\mathbf{G}=\Big[ \mathbf{g_1}~~\mathbf{g_2}~~\mathbf{g_3}~~\mathbf{g_4} \Big]_{2\times 4}$ is the generator matrix of the $(4,2)$-MDS code used to encode the files. Therefore, each server knows the 72 bits $\{\mathbf{u}_i\mathbf{g_n}:1\le i \le36\}$ and $\{\mathbf{v}_i\mathbf{g_n}:1\le i \le36\}$. Upon receiving the query of the form $a_i+b_j$, the $n$-th server responds with $(a_i+b_j)\mathbf{g_n}$, which is a linear combination of the 72 bits stored on the server.

Retrieving the file $U$ is equivalent to retrieving $a_{[1:36]}$ since $S_1$ is a full rank matrix. Note that $a_{[7:36]}$ and $b_{[7:36]}$ could be retrieved since each appears as a query on two different servers and any two of the four vectors $\{\mathbf{g_1},\mathbf{g_2},\mathbf{g_3},\mathbf{g_4}\}$ are linearly independent. Then $b_{[1:6]}$ could be solved from $b_{[7:36]}$ due to the property of $\text{MDS}_{36\times30}$. Therefore, the interferences can be eliminated in the block $\Lambda^{a+b}$ and thus $a_{[1:6]}$ could be retrieved as well.

The scheme is private against any two colluding servers. From the perspective of any two colluding servers, they have received 25 atoms $a_i$, 25 atoms $b_j$ and 5 mixed queries. So altogether they have 30 distinct atoms towards the file $U$ and 30 distinct atoms towards the file $V$. Extract the coefficients of each atom as a vector in $\mathbb{F}_q^{36}$. Recall how we select the random matrices $S_1$, $S_2$ and the $(36,30)$-MDS code. It turns out that the 30 vectors with respect to each file form a random subspace of dimension 30 in $\mathbb{F}_q^{36}$, so the two servers cannot tell any difference between the atoms towards different files and thus the identity of the retrieved file is disguised.

The rate of the scheme above is then $\frac{36\times2}{12\times5\times2+12}=\frac{6}{11}$.

\subsection{$N=4$, $K=2$, $T=2$ and $M=3$}

We now show how the scheme works with one more file than the previous example. Now we have three files $U$, $V$ and $W$. Let each file be of length $432$ and represented in a matrix of size $216\times2$, i.e.,
\begin{equation}
  U=\Bigg(
                  \begin{array}{c}
                    \mathbf{u}_1 \\
                    \vdots \\
                    \mathbf{u}_{216} \\
                  \end{array}
                \Bigg)
  \hspace{3em}
  V=\Bigg(
                  \begin{array}{c}
                    \mathbf{v}_1 \\
                    \vdots \\
                    \mathbf{v}_{216} \\
                  \end{array}
                \Bigg)
  \hspace{3em}
  W=\Bigg(
                  \begin{array}{c}
                    \mathbf{w}_1 \\
                    \vdots \\
                    \mathbf{w}_{216} \\
                  \end{array}
                \Bigg)
\end{equation}
where $\mathbf{u}_i,\mathbf{v}_i,\mathbf{w}_i\in\mathbb{F}_q^{2}$, $1\le i \le 216$. Here $\mathbb{F}_q$ is a sufficiently large finite field. Let $U$ be the desired file.

Similarly as the previous example, we shall construct three lists of atoms $a_{[1:216]}$, $b_{[1:216]}$ and $c_{[1:216]}$, corresponding to one of the files accordingly. These atoms then form the queries to the servers. That is, any query for any server is either a single atom, or a combination of two atoms, or else a mixture of three atoms. We first present the query structure before explaining how to construct the atoms. Similarly as above, the structure is divided into several blocks.

$$\Lambda^{a+b+c}:\Bigg\{\begin{array}{cccc}
  \text{Server I} & \text{Server II} & \text{Server III} & \text{Server IV}\\\hline
  a_{1}+b_{1}+c_{1} & a_{1}+b_{1}+c_{1} & a_{2}+b_{2}+c_{2} & a_{2}+b_{2}+c_{2} \\
  a_{3}+b_{3}+c_{3} & a_{4}+b_{4}+c_{4} & a_{3}+b_{3}+c_{3} & a_{4}+b_{4}+c_{4} \\
  a_{5}+b_{5}+c_{5} & a_{6}+b_{6}+c_{6} & a_{6}+b_{6}+c_{6} & a_{5}+b_{5}+c_{5} \\
\end{array}\Bigg\}\text{,}$$

$$\Lambda^{b+c}_{\lambda}:\Bigg\{\begin{array}{cccc}
  \text{Server I} & \text{Server II} & \text{Server III} & \text{Server IV}\\\hline
  b_{6\lambda+1}+c_{6\lambda+1} & b_{6\lambda+1}+c_{6\lambda+1} & b_{6\lambda+2}+c_{6\lambda+2} & b_{6\lambda+2}+c_{6\lambda+2} \\
  b_{6\lambda+3}+c_{6\lambda+3} & b_{6\lambda+4}+c_{6\lambda+4} & b_{6\lambda+3}+c_{6\lambda+3} & b_{6\lambda+4}+c_{6\lambda+4} \\
  b_{6\lambda+5}+c_{6\lambda+5} & b_{6\lambda+6}+c_{6\lambda+6} & b_{6\lambda+6}+c_{6\lambda+6} & b_{6\lambda+5}+c_{6\lambda+5} \\
\end{array}\Bigg\} \text{ for $\lambda\in\{1,2,3,4,5\}$,}$$

$$\Lambda^{a+b}_{\lambda}:\Bigg\{\begin{array}{cccc}
  \text{Server I} & \text{Server II} & \text{Server III} & \text{Server IV}\\\hline
  a_{6\lambda+1}+b_{6\lambda+31} & a_{6\lambda+1}+b_{6\lambda+31} & a_{6\lambda+2}+b_{6\lambda+32} & a_{6\lambda+2}+b_{6\lambda+32} \\
  a_{6\lambda+3}+b_{6\lambda+33} & a_{6\lambda+4}+b_{6\lambda+34} & a_{6\lambda+3}+b_{6\lambda+33} & a_{6\lambda+4}+b_{6\lambda+34} \\
  a_{6\lambda+5}+b_{6\lambda+35} & a_{6\lambda+6}+b_{6\lambda+36} & a_{6\lambda+6}+b_{6\lambda+36} & a_{6\lambda+5}+b_{6\lambda+35} \\
\end{array}\Bigg\}\text{ for $\lambda\in\{1,2,3,4,5\}$,}$$

$$\Lambda^{a+c}_{\lambda}:\Bigg\{\begin{array}{cccc}
  \text{Server I} & \text{Server II} & \text{Server III} & \text{Server IV}\\\hline
  a_{6\lambda+31}+c_{6\lambda+31} & a_{6\lambda+31}+c_{6\lambda+31} & a_{6\lambda+32}+c_{6\lambda+32} & a_{6\lambda+32}+c_{6\lambda+32} \\
  a_{6\lambda+33}+c_{6\lambda+33} & a_{6\lambda+34}+c_{6\lambda+34} & a_{6\lambda+33}+c_{6\lambda+33} & a_{6\lambda+34}+c_{6\lambda+34} \\
  a_{6\lambda+35}+c_{6\lambda+35} & a_{6\lambda+36}+c_{6\lambda+36} & a_{6\lambda+36}+c_{6\lambda+36} & a_{6\lambda+35}+c_{6\lambda+35} \\
\end{array}\Bigg\}\text{ for $\lambda\in\{1,2,3,4,5\}$,}$$
and finally
$$\Lambda^x_{\eta}:\Bigg\{\begin{array}{cccc}
  \text{Server I} & \text{Server II} & \text{Server III} & \text{Server IV}\\\hline
x_{6\eta+61} & x_{6\eta+61} & x_{6\eta+62} & x_{6\eta+62} \\
x_{6\eta+63} & x_{6\eta+64} & x_{6\eta+63} & x_{6\eta+64} \\
x_{6\eta+65} & x_{6\eta+66} & x_{6\eta+66} & x_{6\eta+65} \\
\end{array}\Bigg\}\text{ for $x\in\{a,b,c\}$ and $\eta\in\{1,\cdots,25\}$.}$$

Independently choose three random matrices $S_1,S_2,S_3\in\mathbb{F}_q^{216\times216}$, uniformly from all the $216\times216$ full rank matrices over $\mathbb{F}_q$. The atoms $a_{[1:216]}$ are built just by setting $a_{[1:216]}=S_1U$.

Suppose there exists a $(36,30)$-MDS code and the transpose of its generator matrix is denoted as $\text{MDS}_{36\times30}$. Select the first 30 rows of $S_2$, denoted as $S_2[(1:30),:]$. Then $b_{[1:36]}$ are built by setting $b_{[1:36]}=\text{MDS}_{36\times30}S_2[(1:30),:]V$. Similarly $c_{[1:36]}=\text{MDS}_{36\times30}S_3[(1:30),:]W$. Note that here we are using the same $(36,30)$-MDS code.

We pause here to explain the functions of the atoms defined so far. Notice that $\{b_i+c_i:7\le i \le 36\}$ could be retrieved since each appears as a query in two different servers. Then due to the property of $\text{MDS}_{36\times30}$, we can solve $\{b_i+c_i:1\le i \le 6\}$ from $\{b_i+c_i:7\le i \le 36\}$. Therefore the interferences can be eliminated in the block $\Lambda^{a+b+c}$ and thus $a_{[1:6]}$ could be retrieved. This is exactly how we make use of the side information provided by a combination of the files $V$ and $W$, to retrieve some messages of the desired file $U$ hidden in the form of a mixture of all three files.

The targets of the other atoms follow the same idea, i.e., we want to make use of the side information provided only by $V$ (respectively, only by $W$) to retrieve some messages of the desired file $U$ hidden in the combination of the files $U$ and $V$ (respectively, the combination of the files $U$ and $W$). This is done in separate parallel steps. Divide the query structure $\{\Lambda^{a+b}_{\lambda}:1\le \lambda \le 5\},\{\Lambda^{a+c}_{\lambda}:1\le \lambda \le 5\},\{\Lambda^x_{\eta}:x\in\{b,c\},1\le \eta \le 25\}$ into the following separate groups:
$$\Gamma^x_{\lambda}=\big\{\Lambda^{a+x}_{\lambda},\{\Lambda^x_{\eta}:5\lambda-4\le \eta \le 5\lambda\} \big\},~x\in\{b,c\},~1\le \lambda \le 5.$$
In each $\Gamma^x_{\lambda}$ we can select the atoms as
$$b_{[6\lambda+31:6\lambda+36]\bigcup[30\lambda+37:30\lambda+66]}=\text{MDS}_{36\times30}S_2[(30\lambda+1:30\lambda+30),:]V$$
or similarly
$$c_{[6\lambda+31:6\lambda+36]\bigcup[30\lambda+37:30\lambda+66]}=\text{MDS}_{36\times30}S_3[(30\lambda+1:30\lambda+30),:]W.$$

Retrieving the file $U$ is equivalent to retrieving $a_{[1:216]}$ since $S_1$ is a full rank matrix. Retrieving $a_{[67:216]}$ is straightforward and we have explained how to retrieve $a_{[1:6]}$. In each $\Gamma^b_{\lambda}$ or $\Gamma^c_{\lambda}$ we have the atoms $a_{[6\lambda+1:6\lambda+6]}$ or $a_{[6\lambda+31:6\lambda+36]}$ accompanied by the interferences $b_{[6\lambda+31:6\lambda+36]}$ or $c_{[6\lambda+31:6\lambda+36]}$. Due to the MDS property we can solve these interferences via $b_{[30\lambda+37:30\lambda+66]}$ or $c_{[30\lambda+37:30\lambda+66]}$. Once eliminating these interferences, we are able to retrieve $a_{[7:66]}$.

The scheme is private against any two colluding servers. From the perspective of any two colluding servers, they have received atoms towards a single file 125 times each, queries towards a combination of two files 25 times each and finally 5 queries towards a mixture of all the three files. So altogether the number of atoms towards each of the three files is $125+25\times2+5=180$. Extract the coefficients of each atom as a vector in $\mathbb{F}_q^{216}$. Recall how we select the random matrices $S_1$, $S_2$, $S_3$ and the $(36,30)$-MDS code. It turns out that the 180 vectors with respect to each file form a random subspace of dimension 180 in $\mathbb{F}_q^{216}$, so the two servers cannot tell any difference among the atoms towards different files and thus the identity of the retrieved file is disguised.

The rate of the scheme above is then $\frac{216\times2}{12\times25\times3+12\times5\times3+12}=\frac{36}{91}$.

\subsection{The general framework}

From the illustrations of the two examples above, we proceed to explain the general framework of our PIR scheme. Denote the files as $W^{[1]},W^{[2]},\dots,W^{[M]}$. Each file is of length $LK$ and represented in a matrix of size $L\times K$ over $\mathbb{F}_q$, where $L$ is a constant to be computed later. $\mathbb{F}_q$ is a sufficiently large field that allows the existence of an MDS code used in the scheme later. Each file, say $W^{[m]}$, consists of $L$ rows, $\mathbf{w}^{[m]}_1,\mathbf{w}^{[m]}_2,\dots,\mathbf{w}^{[m]}_L$. For each file $W^{[m]}$ we will build a list of $L$ atoms, where each atom is a linear combination of $\{\mathbf{w}^{[m]}_1,\mathbf{w}^{[m]}_2,\dots,\mathbf{w}^{[m]}_L\}$. Let $W^{[1]}$ be the desired file. Constructing the PIR scheme contains the following steps.

$\bullet$ Step 1: Independently choose $M$ random matrices, $S_1,\dots,S_M$, uniformly from all the $L\times L$ full rank matrices over $\mathbb{F}_q$. Then the atoms for $W^{[1]}$ are just built by $S_1W^{[1]}$.

$\bullet$ Step 2: Let $\alpha$ and $\beta$ be the smallest positive integers satisfying
\begin{equation} \label{equation}
\alpha{{N}\choose{K}}=(\alpha+\beta)\Bigg({{N}\choose{K}}-{{N-T}\choose{K}}\Bigg).
\end{equation}
Assume the existence of an $\big((\alpha+\beta){{N}\choose{K}},\alpha{{N}\choose{K}}\big)$-MDS code. The transpose of its generator matrix is denoted as $\text{MDS}_{(\alpha+\beta){{N}\choose{K}}\times\alpha{{N}\choose{K}}}$.

$\bullet$ Step 3: We need an {\it assisting array} of size ${{N-1}\choose{K-1}}\times N$ consisting of ${{N}\choose{K}}$ symbols. Each symbol appears $K$ times and every $K$ columns share a common symbol. For example, when $N=4$, $K=2$ the array is of the form
$$\Bigg\{\begin{array}{cccc}
  1 & 1 & 2 & 2 \\
  3 & 4 & 3 & 4 \\
  5 & 6 & 6 & 5
\end{array}\Bigg\}.$$

$\bullet$ Step 4: \emph{[Construction of the query structure]} The query structure is divided into several blocks, where each block is labelled by $\mathcal{F}\subseteq\{W^{[1]},W^{[2]},\dots,W^{[M]}\}$, a subset of files. In a block labelled by $\mathcal{F}$, the queries are of the same form, i.e., every query is a mixture of $|\mathcal{F}|$ atoms related to the files in $\mathcal{F}$. We set each block in an ``isomorphic" form with the assisting array, i.e., every $K$ servers share a common query. For example, each block of queries in the previous examples is isomorphic with the assisting array above.

We further call a block labelled by $\mathcal{F}$ a {\it $t$-block} if $|\mathcal{F}|=t$. For any $\mathcal{F}\subseteq\{W^{[1]},W^{[2]},\dots,W^{[M]}\}$, $|\mathcal{F}|=t$, we require that the number of $t$-blocks labelled by $\mathcal{F}$ is $\alpha^{M-t}\beta^{t-1}$. For example, in Subsection B, $\alpha=5$ and $\beta=1$. So we have one 3-block labelled by $a+b+c$ (equivalently, the set $\{U,V,W\}$), five 2-blocks for each label $a+b,a+c,b+c$ (equivalently, the sets $\{U,V\}$, $\{U,W\}$ and $\{V,W\}$) and twenty-five 1-blocks for each label $a,b,c$ (equivalently, the singleton set $\{U\}$, $\{V\}$ and $\{W\}$).

$\bullet$ Step 5: \emph{[Dividing the query structure into groups]} Now let $\mathcal{F}$ denote a nonempty set of files not containing the desired file $W^{[1]}$. There are totally $\alpha^{M-|\mathcal{F}|}\beta^{|\mathcal{F}|-1}$ blocks labelled by $\mathcal{F}$ and $\alpha^{M-|\mathcal{F}|-1}\beta^{|\mathcal{F}|}$ blocks labelled by $\mathcal{F}\bigcup\{W^{[1]}\}$. Divide these blocks into $\alpha^{M-|\mathcal{F}|-1}\beta^{|\mathcal{F}|-1}$ groups, denoted by $\Gamma^{\mathcal{F}}_{\lambda}$, $1\le\lambda\le \alpha^{M-|\mathcal{F}|-1}\beta^{|\mathcal{F}|-1}$, where each group consists of $\alpha$ blocks labelled by $\mathcal{F}$ and $\beta$ blocks labelled by $\mathcal{F}\bigcup\{W^{[1]}\}$. Repeat this process for every nonempty set of files not containing the desired file $W^{[1]}$ and thus the whole query structure is divided into groups.

$\bullet$ Step 6: \emph{[Constructing the atoms for each group]} For each group constructed with respect to $\mathcal{F}$ and any file $W^{[m]}$ in $\mathcal{F}$, the number of atoms towards $W^{[m]}$ within the blocks in this group is $(\alpha+\beta){{N}\choose{K}}$, among which $\alpha{{N}\choose{K}}$ atoms appear in blocks labelled by $\mathcal{F}$ and the other $\beta{{N}\choose{K}}$ atoms appear as interferences in blocks labelled by $\mathcal{F}\bigcup\{W^{[1]}\}$. Let these $(\alpha+\beta){{N}\choose{K}}$ atoms for $W^{[m]}$ be built by $\text{MDS}_{(\alpha+\beta){{N}\choose{K}}\times\alpha{{N}\choose{K}}}S'_mW^{[m]}$, where $S'_m$ denotes some $\alpha{{N}\choose{K}}$ rows in the matrix $S_m$.

Repeat this process for all the groups. Note that we shall come across the same file $W^{[M]}$ several times. We have ${{M-2}\choose{|\mathcal{F}|-1}}$ choices for a subset $\mathcal{F}$ of size $|\mathcal{F}|$, containing $W^{[M]}$ but without $W^{[1]}$. With respect to any such $\mathcal{F}$ we have $\alpha^{M-|\mathcal{F}|-1}\beta^{|\mathcal{F}|-1}$ groups. So the exact number of times we come across $W^{[M]}$ can be computed as
\begin{equation}
  \sum_{|\mathcal{F}|=1}^{M-1} \alpha^{M-|\mathcal{F}|-1}\beta^{|\mathcal{F}|-1}{{M-2}\choose{|\mathcal{F}|-1}}=(\alpha+\beta)^{M-2}.
\end{equation}

Every time we come across the same file $W^{[M]}$, we have to ensure that the rows we select from $S_m$ are non-intersecting. This is guaranteed as long as
\begin{equation} \label{Inequality}
  (\alpha+\beta)^{M-2}\alpha{{N}\choose{K}}\le L.
\end{equation}

$\bullet$ Step 0: We leave the determination of $L$ here since its value is deduced based on all the steps above. However, we call this step by Step 0 since the value of $L$ should be used from the very beginning. $L$ is exactly the number of distinct atoms towards the desired file. So $L$ can be computed as
\begin{equation}
  L={{N}\choose{K}}\sum_{|\mathcal{F}|=1}^{M} \alpha^{M-|\mathcal{F}|}\beta^{|\mathcal{F}|-1}{{M-1}\choose{|\mathcal{F}|-1}}={{N}\choose{K}}(\alpha+\beta)^{M-1}.
\end{equation}

One can see that the value of $L$ will guarantee the correctness of the inequality (\ref{Inequality}).

\subsection{Analysis of the PIR scheme}

The underlying $(N,K)$-MDS code used in the distributed storage system guarantees that any $K$ encoding vectors are linearly independent. Thus each query in the scheme can be retrieved since the same query appears in $K$ distinct servers. For each group constructed above with respect to $\mathcal{F}$, based on the property of the $\big((\alpha+\beta){{N}\choose{K}},\alpha{{N}\choose{K}}\big)$-MDS code, those $\alpha{{N}\choose{K}}$ queries in the blocks labelled by $\mathcal{F}$ will help us eliminate the interferences of the $\beta{{N}\choose{K}}$ queries in the blocks labelled by $\mathcal{F}\bigcup\{W^{[1]}\}$, and thus the $\beta{{N}\choose{K}}$ atoms towards $W^{[1]}$ are retrieved. The retrieval for those atoms towards $W^{[1]}$ not interfered are straightforward. As a result, all the atoms $a_{[1:L]}$ are successfully retrieved, which is equivalent to the retrieval of the desired file $W^{[1]}$, since $S_1$ is of full rank.

The scheme is private against any $T$ colluding servers. From the perspective of any $T$ colluding servers, for each group constructed above with respect to $\mathcal{F}$, they shall receive $(\alpha+\beta)\big({{N}\choose{K}}-{{N-T}\choose{K}}\big)$ atoms towards each file, say $W^{[m]}\in\mathcal{F}$. By the equality (\ref{equation}) this number is exactly $\alpha{{N}\choose{K}}$. Extract the coefficients of the atom as a vector in $\mathbb{F}_q^{L}$. Recall how we build the atoms in this group and the property of the $\big((\alpha+\beta){{N}\choose{K}},\alpha{{N}\choose{K}}\big)$-MDS code. One can see that the vectors will form a random subspace of dimension $\alpha{{N}\choose{K}}$ in $\mathbb{F}_q^{L}$. Moreover, recall that the atoms towards $W^{[m]}$ in different groups are made from non-intersecting rows of $S_m$ and $S_m$ is of full rank. Thus in the whole query structure, the vectors of all the atoms towards $W^{[m]}$ will form a random subspace of dimension $(\alpha+\beta)^{M-2}\alpha{{N}\choose{K}}$ in $\mathbb{F}_q^{L}$. This property also trivially holds for the file $W^{[1]}$. So these $T$ colluding servers cannot tell any difference among the atoms towards different files and thus the identity of the retrieved file is disguised.

Finally we compute the rate of our PIR scheme. The retrieved file is of length $LK$. The download cost in any block is $K{N\choose K}$. So the rate can be computed as
$$\frac{LK}{K{N\choose K}\sum_{|\mathcal{F}|=1}^{M} \alpha^{M-|\mathcal{F}|}\beta^{|\mathcal{F}|-1}{{M}\choose{|\mathcal{F}|}}}=\frac{L}{{N\choose K}\beta^{-1}\big((\alpha+\beta)^M-\alpha^M\big)}=\frac{1}{1+\frac{\alpha}{\alpha+\beta}+\cdots+(\frac{\alpha}{\alpha+\beta})^{M-1}}.$$

A final remark is that our scheme, like most existing schemes, only works for the case $T+K\le N$ since otherwise the equality (\ref{equation}) does not make sense. Another way to explain this limitation is as follows. We build the queries according to the assisting array. If $T+K>N$ then every $T$ servers will know all the queries and all the atoms and thus no side information can be used.

To sum up, our main result is as follows.

\begin{theorem}
  When $T+K\le N$, there exists an $(N,K,T;M)$-PIR scheme with rate $(1+R+R^2+\cdots+R^{M-1})^{-1}$, where $R=1-\frac{{{N-T}\choose K}}{{N\choose K}}$.
\end{theorem}

\section{A comparison with known PIR schemes}\label{SecComparison}

In this section we briefly discuss some other known PIR schemes for MDS coded databases with colluding servers and make some comparisons.

First of all, for the degenerating cases, i.e., either $K=1$ or $T=1$, the rate of our scheme agrees with previous results in \cite{BanawanCoded,Sun,SunColluded}. For the non-degenerating cases, one can see that our rate is strictly smaller than the conjectured capacity in (\ref{FGHKConjecture}). So our result offers a lower bound of the capacity and does not provide any counterexamples to the conjecture by Freij-Hollanti et al.

The scheme by Freij-Hollanti et al. in \cite{Freij-Hollanti} has rate $\frac{N-K-T+1}{N}$. One drawback of the scheme is that the underlying MDS code in the distributed storage system should satisfy some certain properties. Regardless of this drawback, the scheme performs very well when the number of files $M$ is relative large and performs bad when $M$ is relatively small. One can see that given $N$, $K$ and $T$, the rate of our scheme is a strictly decreasing function of $M$ and the limitation is $\frac{{{N-T}\choose K}}{{N\choose K}}$. The rate of the scheme in \cite{Freij-Hollanti} is a constant independent of $M$ and $\frac{N-K-T+1}{N}$ is larger than $\frac{{{N-T}\choose K}}{{N\choose K}}$. So there will be a threshold $M(N,K,T)$. Our scheme is better when the number of files is less than $M(N,K,T)$ and the scheme in \cite{Freij-Hollanti} is better otherwise. As an example, consider the case with parameters $N=30$, $K=20$ and $T=10$. $R_1$ represents the rate for our scheme, shown in the red curve with circles. $R_2$ represents the rate of the scheme by Freij-Hollanti et al., shown by the green straight line. $C$ denotes the conjectured capacity by Freij-Hollanti et al. in (\ref{FGHKConjecture}), shown in the blue curve with triangles. Then the threshold is $M(30,20,10)=30$ (when the number of files is exactly 30, our scheme has a better rate with only a slight difference of $1.6\times10^{-8}$).

\begin{figure}[h]
\centering
\includegraphics[height=8cm,width=17cm]{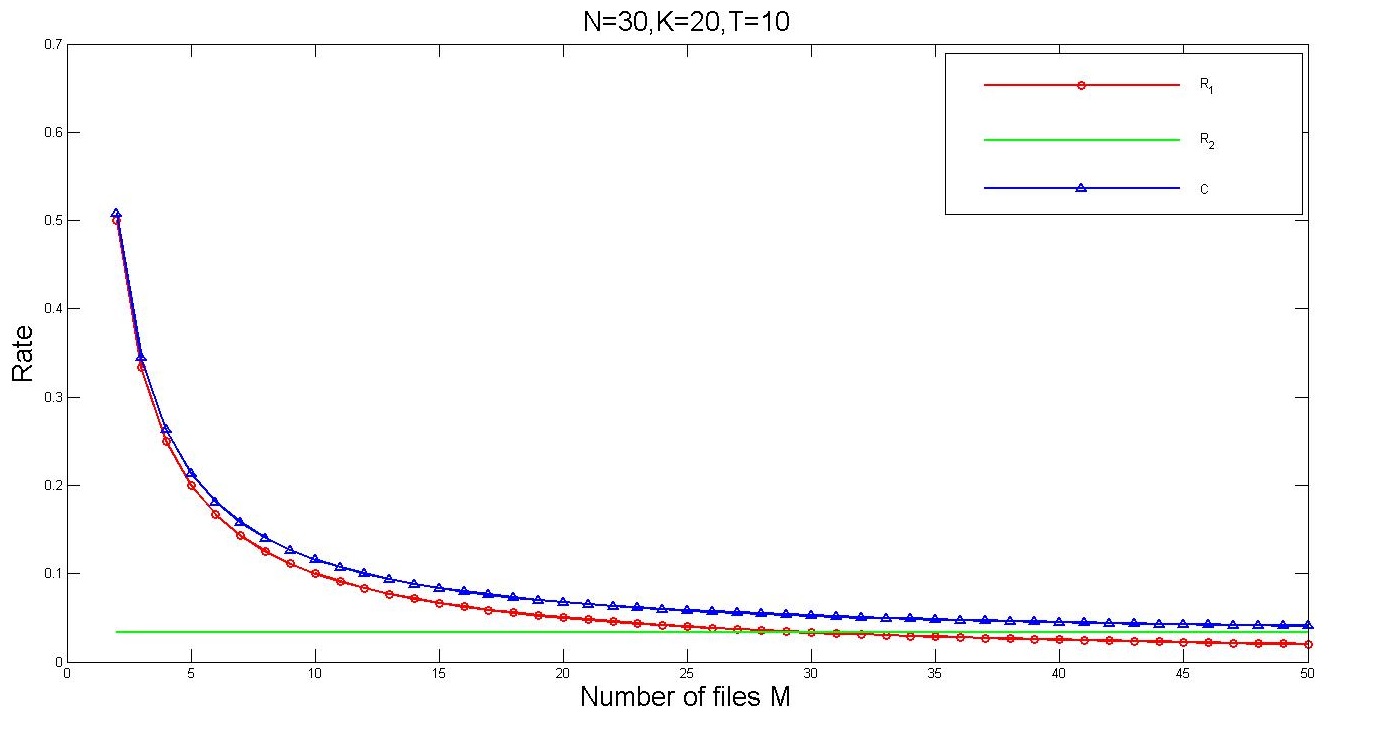}
\end{figure}

The recent work by Sun and Jafar in \cite{SunNew} is remarkable since it allows a coding strategy for the servers before responding. This has been a blind spot for all previous works and also the current draft. Simply speaking, the idea in \cite{SunNew} is to make the query space (spaces spanned by the query vectors) larger for the desired file and smaller for the other files. The query space is sent to each server individually (unlike in our scheme which contains mixed queries). The servers perform some coding procedures to produce some feedbacks which is a mixture of several files and thus reduce the download cost. By analyzing the case with parameters $N=4$, $K=2$, $T=2$ and $M=2$, the scheme in \cite{SunNew} has rate $\frac{3}{5}$, which is larger than the conjectured optimal value $\frac{4}{7}$ and consequently larger than $\frac{6}{11}$ in our first example. Another important advantage of the scheme by Sun and Jafar is that, as far as we know, it is the first scheme suitable for the case $T+K>N$.

However, a naive generalization of the scheme in \cite{SunNew} seems to be not satisfying. Suppose the dimension of the query space for the desired file is $D$ and the dimension of the query space for any other file is $d$. Then the scheme based on the same idea will only have a rate $\frac{D}{D+d(M-1)}$, which tends to zero very fast with the growing of $M$. So the scheme in \cite{SunNew} may only have very good performance when the number of files is very small. Maybe we have not captured the essence of the scheme in \cite{SunNew} and its generalization will be of great interest.

Finally we show a comparison of all the schemes with the running example throughout the draft, $N=4$, $K=2$ and $T=2$. The meanings of $R_1$, $R_2$ and $C$ are as mentioned above. $R_3$ represents the rate of the new scheme by Sun and Jafar (or more precisely, its generalization based on our understanding), shown in the black curve with squares.

\begin{figure}[h]
\centering
\includegraphics[height=8cm,width=17cm]{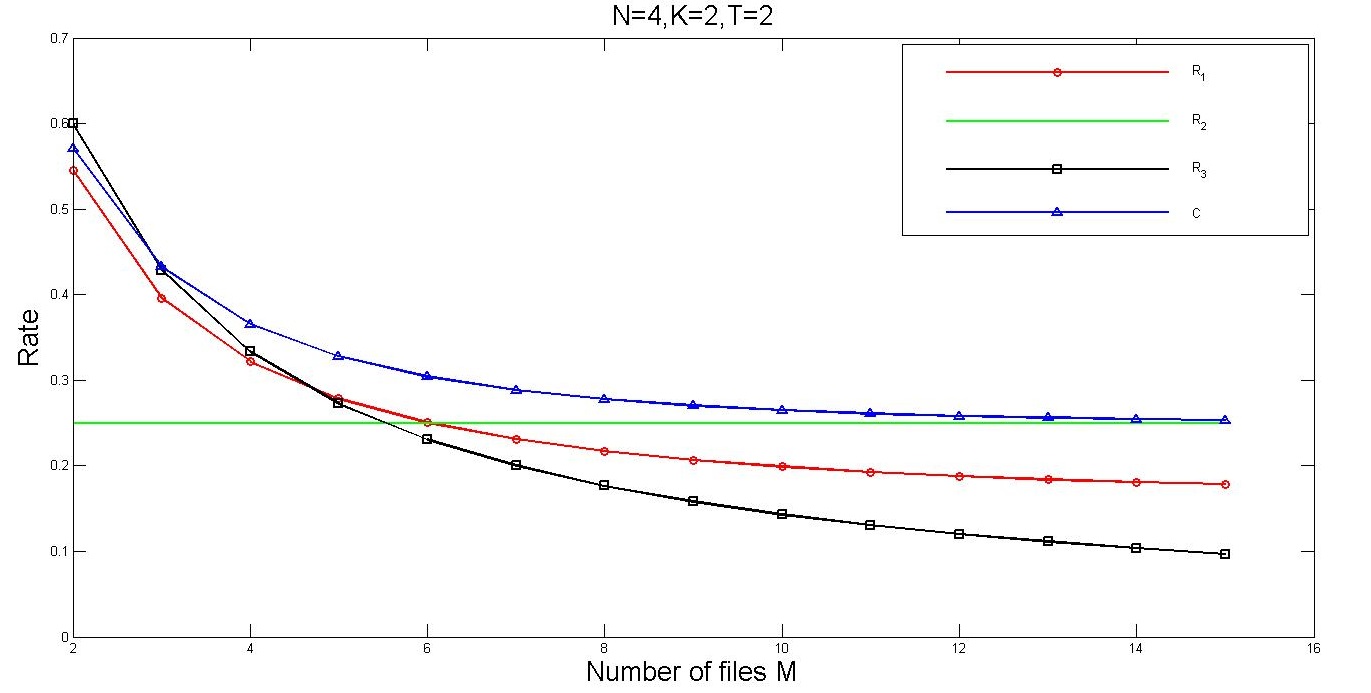}
\end{figure}

\begin{center}
\begin{tabular}{|c|c|c|c|c|c|}
  \hline
  ~ & $R_1$ & $R_2$ & $R_3$ & $C$ & Order\\\hline
  $M=2$ & 0.5454 & 0.2500 & 0.6000 & 0.5714 & $R_3>C>R_1>R_2$\\\hline
  $M=3$ & 0.3956 & 0.2500 & 0.4286 & 0.4324 & $C>R_3>R_1>R_2$\\\hline
  $M=4$ & 0.3219 & 0.2500 & 0.3333 & 0.3657 & $C>R_3>R_1>R_2$\\\hline
  $M=5$ & 0.2786 & 0.2500 & 0.2727 & 0.3278 & $C>R_1>R_3>R_2$\\\hline
  $M=6$ & 0.2506 & 0.2500 & 0.2308 & 0.3041 & $C>R_1>R_2>R_3$\\\hline
  $M=7$ & 0.2312 & 0.2500 & 0.2000 & 0.2885 & $C>R_2>R_1>R_3$\\
  \hline
\end{tabular}
\end{center}

When $M=2$, $R_3$ is the largest, which results in a counterexample to the conjecture by Freij-Hollanti et al. $R_3$ still performs well when the file number is $3$ or $4$, but no longer violates the conjecture. When $M=5$ or $M=6$, our scheme becomes the best. When $M\ge7$, the order is then always $C>R_2>R_1>R_3$.

\section{Conclusion}\label{SecConclusion}

We have constructed a general private information retrieval scheme for $(N,K)$-MDS coded database with arbitrary $T$-colluding servers. The rate of our scheme is $(1+R+R^2+\cdots+R^{M-1})^{-1}$, where $R=1-\frac{{{N-T}\choose K}}{{N\choose K}}$. Our scheme performs better than existing schemes for a certain range of parameters. One more advantage of our scheme is that it works for any underlying MDS code used in the distributed storage system. In general, determining the exact PIR capacity when $K\ge2$ and $T\ge2$ is far from solved. New schemes increasing the rate (even only for a small range of parameters) and new upper bounds for the capacity will be of great interest.

In the end we briefly introduce various other models on PIR. The symmetric PIR problem is considered in \cite{SunSymmetric} and \cite{Wang}, where a further constraint is that the user should know nothing about any non-retrieved file. \cite{BanawanMultimessage} considers the model for retrieving $P\ge 2$ files and shows that one can do better than the trivial approach of executing $P$ independent PIR schemes. \cite{SunMultiround} discusses multi-round PIR schemes, where the queries and feedbacks are made in several rounds and the user may adjust his queries according to the feedbacks from previous rounds. \cite{TajeddineNew} considers arbitrary collusion patterns instead of the original ``$T$ out of the $N$ servers may collude" model. \cite{Kumar} considers the model replacing the underlying MDS code by some non-MDS code.


\begin{thebibliography}{99}
\bibitem{Asi}
H. Asi and E. Yaakobi, Nearly optimal constructions of PIR and batch codes, arXiv preprint arXiv:1701.07206, 2017.
\bibitem{Augot}
D. Augot, F. Levy-dit-Vehel and A. Shikfa, A storage-efficient and robust Private Information Retrieval Scheme allowing few servers, in Proceedings of Cryptology and Network Security (CANS), pp. 222-239, 2014.
\bibitem{BanawanCoded}
K. Banawan and S. Ulukus, The capacity of private information retrieval from coded databases, arXiv preprint arXiv:1609.08138, 2016.
\bibitem{BanawanMultimessage}
K. Banawan and S. Ulukus, Multi-message private information retrieval: capacity results and near-optimal schemes, arXiv preprint arXiv:1702.01739, 2017.
\bibitem{Beimel}
A. Beimel, Y. Ishai, E. Kushilevitz and J.-F. Raymond, Breaking the $O(n^{1/(2k-1)})$ barrier for information-theoretic private information retrieval, in Proceedings of the Annual Symposium on Foundations of Computer Science (FOCS), pp. 261-270, 2002.
\bibitem{BlackburnPIRcodes}
S. Blackburn and T. Etzion, PIR array codes with optimal PIR rate, arXiv preprint arXiv:1607.00235, 2016.
\bibitem{Blackburn}
S. Blackburn, T. Etzion and M. Paterson, PIR schemes with small download complexity and low storage requirements, arXiv preprint arXiv:1609.07027, 2016.
\bibitem{Chan}
T. Chan, S. Ho and H. Yamamoto, Private information retrieval for coded storage, in Proceedings of IEEE International Symposium on Information Theory (ISIT), pp. 2842-2846, 2015.
\bibitem{Chor1}
B. Chor, O. Goldreich, E. Kushilevitz and M. Sudan, Private information retrieval, in Proceedings of the Annual Symposium on Foundations of Computer Science (FOCS), pp. 41-50, 1995.
\bibitem{Chor2}
B. Chor, E. Kushilevitz, O. Goldreich and M. Sudan, Private information retrieval, Journal of the ACM, vol. 45, no. 6, pp. 965-981, 1998.
\bibitem{Dvir}
Z. Dvir and S. Gopi, 2-server PIR with sub-polynomial communication, Journal of the ACM, vol. 63, no. 4, article 39, 2016.
\bibitem{Efremenko}
K. Efremenko, 3-query locally decodable codes of subexponential length, SIAM Journal on Computing, vol. 41, no. 6, pp. 1694-1703, 2012.
\bibitem{FazeliConfversion}
A. Fazeli, A. Vardy and E. Yaakobi, Codes for Distributed PIR with Low Storage Overhead, in Proceedings of IEEE International Symposium on Information Theory (ISIT), pp. 2852-2856, 2015.
\bibitem{FazeliFullversion}
A. Fazeli, A. Vardy and E. Yaakobi, PIR with low storage overhead: coding instead of replication, arXiv preprint arXiv:1505.06241, 2015.
\bibitem{Freij-Hollanti}
R. Freij-Hollanti, O. Gnilke, C. Hollanti and D. Karpuk, Private information retrieval from coded databases with colluding servers, arXiv preprint arXiv:1611.02062, 2016.
\bibitem{Kumar}
S. Kumar, E. Rosnes and A. G. i Amat, Private information retrieval in distributed storage systems using an arbitrary linear code, arXiv preprint arXiv:1612.07084, 2016.
\bibitem{Rao}
S. Rao and A. Vardy, Lower bound on the redundancy of PIR codes, arXiv preprint arXiv:1605.01869, 2016.
\bibitem{Shah}
N. B. Shah, K. V. Rashmi and K. Ramchandran, One extra bit of download ensures perfectly private information retrieval, in Proceedings of IEEE International Symposium on Information Theory (ISIT), pp. 856-860, 2014.
\bibitem{SunMultiround}
H. Sun and S. Jafar, Multiround Private Information Retrieval: Capacity and Storage Overhead, arXiv preprint arXiv:1611.02257, 2016.
\bibitem{SunNew}
H. Sun and S. Jafar, Private information retrieval from MDS coded data with colluding servers: settling a conjecture by Freij-Hollanti et al., arXiv preprint arXiv:1701.07807, 2017.
\bibitem{Sun}
H. Sun and S. Jafar, The capacity of private information retrieval, arXiv preprint arXiv:1602.09134, 2016.
\bibitem{SunColluded}
H. Sun and S. Jafar, The capacity of robust private information retrieval with colluding databases, arXiv preprint arXiv:1605.00635, 2016.
\bibitem{SunSymmetric}
H. Sun and S. Jafar, The capacity of symmetric private information retrieval, arXiv preprint arXiv: 1606.08828, 2016.
\bibitem{TajeddineConfversion}
R. Tajeddine and S. E. Rouayheb, Private information retrieval from MDS coded data in distributed storage systems, in Proceedings of IEEE International Symposium on Information Theory (ISIT), pp. 1411-1415, 2016.
\bibitem{TajeddineFullversion}
R. Tajeddine and S. E. Rouayheb, Private information retrieval from MDS coded data in distributed storage systems, arXiv preprint arXiv:1602.01458, 2016.
\bibitem{TajeddineNew}
R. Tajeddine, O. Gnilke, D. Karpuk, R. Freij-Hollanti, C. Hollanti and S. E. Rouayheb, Private information retrieval schemes for coded data with arbitrary collusion patterns, arXiv preprint arXiv:1701.07636, 2017.
\bibitem{Vajha}
M. Vajha, V. Ramkumar and P. V. Kumar, Binary, shortened projective Reed Muller codes for coded private information retrieval, arXiv preprint arXiv:1702.05074, 2017.
\bibitem{Wang}
Q. Wang and M. Skoglund, Symmetric private information retrieval for MDS coded distributed storage, arXiv:1610.04530, 2016.
\bibitem{YekhaninSurvey}
S. Yekhanin, Private information retrieval, Communications of the ACM, vol. 53, no. 4, pp. 68-73, 2010.
\bibitem{Yekhanin}
S. Yekhanin, Towards 3-query locally decodable codes of subexponential length, Journal of the ACM, vol. 55, no. 1, article 1, 2008.
\bibitem{Zhang}
Y. Zhang, X. Wang, H. Wei and G. Ge, On private information retrieval array codes, arXiv preprint arXiv:1609.09167, 2016.

\end{thebibliography}
\end{document}